%% file: ms.tex
\documentclass[12pt,preprint]{aastex}

\received{}
\accepted{}

\slugcomment{Submitted to ApJ}

\newcommand{\rhk}{R$^{\prime}_{\mathrm{HK}}$}
\newcommand{\ms}{\mbox{m\,s$^{-1}$}}
\newcommand{\kms}{\mbox{km\,s$^{-1}$}}

\newcommand{\mjup}{M$_{\rm J}$}

\newcommand{\msini}{$M \sin i$}

\newcommand{\vsini}{$v\sin i$}
\newcommand{\Mv}{M$_{\rm v}~$}
\newcommand{\simgt}{\lower.5ex\hbox{$\; \buildrel > \over \sim \;$}}
\newcommand{\simlt}{\lower.5ex\hbox{$\; \buildrel < \over \sim \;$}}
\newcommand{\poneper}{3.33\,yr}

\newcommand{\pfourbper}{7.94\,yr}
\newcommand{\poneecc}{0.44}

\newcommand{\pthreeecc}{0.69}

\newcommand{\pfouraecc}{0.47}
\newcommand{\pfourbecc}{0.87}

\newcommand{\ponemass}{1.5}

\newcommand{\pthreemass}{1.7}
\newcommand{\pfouramass}{1.8}
\newcommand{\pfourbmass}{18.4}
\newcommand{\poneamp}{27.5}

\newcommand{\pthreeamp}{43.3}
\newcommand{\pfouraamp}{50.4}
\newcommand{\pfourbamp}{466.5}
\newcommand{\ponechisq}{5.11}

\newcommand{\pthreechisq}{15.22}
\newcommand{\pfourchisq}{1.73}
\newcommand{\ponenobs}{34}

\newcommand{\pthreenobs}{28}
\newcommand{\pfournobs}{28}
\newcommand{\ponesep}{2.4}

\newcommand{\pthreesep}{2.0}
\newcommand{\pfourasep}{1.2}
\newcommand{\pfourbsep}{4.3}

\shortauthors{O'Toole {\it et~al.\/}}
\shorttitle{Planets Around G Dwarfs}
\begin{document}

\title{New Planets around three G Dwarfs$~^{1}$}

\author{Simon J. O'Toole\altaffilmark{2}, 
R. Paul Butler\altaffilmark{3},
C. G. Tinney\altaffilmark{2},
Hugh R. A. Jones\altaffilmark{4},
Geoffrey W. Marcy\altaffilmark{5,7},
Brad Carter\altaffilmark{6},
Chris McCarthy\altaffilmark{7},
Jeremy Bailey\altaffilmark{8},
Alan J. Penny\altaffilmark{9},
Kevin Apps\altaffilmark{10},
Debra Fischer\altaffilmark{7,5}
}

\altaffiltext{1}{Based on observations obtained at the
Anglo--Australian Telescope, Siding Spring, Australia.}

\altaffiltext{2}{Anglo--Australian Observatory, P.O. Box 296,
Epping, NSW 1710, Australia}

\altaffiltext{3}{Department of Terrestrial Magnetism, Carnegie
Institution of Washington, 5241 Broad Branch Road NW,
Washington DC, USA 20015-1305}

\altaffiltext{4}{Centre for Astrophysics Research, University
of Hertfordshire, Hatfield, AL 10 9AB, UK}

\altaffiltext{5}{Department of Astronomy, University of California,
Berkeley, CA USA 94720}

\altaffiltext{6}{Faculty of Sciences, University of Southern
Queensland, Toowoomba, Queensland 4350, Australia}

\altaffiltext{7}{Department of Physics and Astronomy,
San Francisco State University, San Francisco, CA, USA 94132}

\altaffiltext{8}{Australian Centre for Astrobiology,
Macquarie University, Sydney, NSW 2109, Australia}

\altaffiltext{9}{University of St Andrews, School of Physics
and Astronomy, North Haugh, St Andrews, UK}

\altaffiltext{10}{Physics and Astronomy, University of Sussex,
Falmer, Brighton BN1 9QJ, UK}

\begin{abstract}
Doppler velocity measurements from the Anglo-Australian Planet Search
reveal planetary mass companions to HD\,23127, HD\,159868,
and a possible second planetary companion to HD\,154857.  These stars are
all G dwarfs. The companions are all in eccentric orbits with periods
ranging from 1.2 to $>9.3$\,yr, minimum (\msini) masses ranging from 1.5 to
$>4.5$\,\mjup, and semimajor axes between 1 and $>4.5$\,AU.  The
orbital parameters are updated for the inner planet to HD\,154857,
while continued monitoring of the outer companion is required to
confirm its planet status.  
\end{abstract}


\section{Introduction}
\label{intro}

As extrasolar planetary research enters its second decade, the field
continues to be driven by precision Doppler surveys.  Roughly 95\%
of all known planets come from such surveys, including
all of the $\sim$180 extrasolar planets\footnote{cf. list of exoplanets orbiting
nearby stars at www.exoplanets.org at 1 October 2006} orbiting
stars within 100\,pc (Butler et al.\ 2006).

The  most promising new methods for the detection and study
of nearby exoplanets are the next generation techniques
of interferometric astrometry and direct imaging. However,
these have proven harder to develop
and implement than expected a decade ago.  Over the same period
Doppler programs have steadily improved their precision from 10 to 1\,\ms.  
At this level, precision Doppler surveys will remain the dominant
detection technique for exoplanets orbiting nearby stars for the
foreseeable future.

The Anglo-Australian Planet Search (AAPS) began observing 200
stars in January 1998.  In 2002 the program expanded from 20 to 32
nights per year.  In response two changes were made: 60 new stars
were added, and observations with signal-to-noise ratios of at least
200 became the goal for all survey stars.
The AAT target list has been published in Jones et al.\ (2002).

A total of 26 planets have emerged from the AAT program.
The full set of AAT Doppler velocity measurements for these stars are
included in the ``Catalog of Nearby Exoplanets'' (Butler et al.\ 2006).
These AAT velocities have
already yielded a new 300 day planet orbiting HD\,160691 (Gozdiewski
et al.\ 2006), which has been subsequently confirmed with HARPS data
(Pepe et al.\ 2006). In this paper three new AAT planets are announced.

\section{Observations}

Precision Doppler velocity measurements  are
made at the 3.9-m Anglo-Australian
Telescope (AAT) with the UCLES echelle spectrograph (Diego et al.\ 1990). 
A 1\arcsec\ slit yields $\lambda/\Delta\lambda \sim 50000$
spectra that span the wavelength range from 4820--8550\,\AA.  An
iodine absorption cell (Marcy \& Butler 1992) provides wavelength
calibration from 5000 to 6100\,\AA.  The spectral point-spread
function is derived
from the detailed shapes of the embedded iodine lines
(Valenti et al.\ 1995).  The precision Doppler analysis is
carried out with an updated version of the technique
outlined by Butler et al.\ (1996).
The long-term underlying systematic precision of the AAPS 
as demonstrated by stable stars 
is 3\,\ms (see e.g. Figures 1-4 of Butler et al.\ 2001, Figure 1
of McCarthy et al.\ 2004, Figure 1 of Tinney et al.\ 2005). 
Only the Doppler program at Keck has demonstrated a similar level of
precision on time scales of many years (see Figures in Vogt et
al.\ 2000, Butler et al.\ 2004, Marcy et al.\ 2005, Vogt et al.\ 2005,
Rivera et al.\ 2005, and Butler et al.\ 2006). 

\section{Three New Planets}
\label{sec:4new}
The physical parameters for the planet-bearing stars reported in
this paper are listed in Table \ref{candidprop}.    
All three of these stars are classified as G dwarfs in the Michigan
Catalog (Houk \& Cowley 1975, 1978, 1982). The stellar
distances are from HIPPARCOS (Perryman et al.\ 1997). The estimates of
stellar activity (\rhk) are from Jenkins et al.\ (2006).  The
metallicity and \vsini\ measurements are from Valenti \& Fischer
(2005) based on spectral synthesis matched to high-resolution
($\lambda/\Delta\lambda \sim 66,000$) iodine-free ``template''
spectra, taken with a 0.75\arcsec\ slit.  The estimates of stellar
mass and age are from the analysis of Takeda et al.\ (2006).  The
estimated intrinsic stellar Doppler jitter is from our latest upgrade
to the calibration of Wright (2005).

The absolute magnitudes of these stars range
from \Mv $=$ 3.8 to \Mv $=$ 3.0, suggesting they
have all begun to evolve off the main sequence, consistent with
their estimated stellar ages (5.6 to 8 Gyr), and with the
measured level of chromospheric activity and \vsini\ velocities.

\subsection{HD\,159868}

This star has been observed \pthreenobs\ times since being
added to the AAPS program when it was expanded in 2002. These data are
listed in Table \ref{vel159868} and shown in Figure \ref{fig4}. 
The root-mean-square (RMS) of these velocities 
about the mean is 29\,\ms, exceeding that expected from the photon-counting
internal measurement precision and stellar jitter. (The median value of the 
internal measurement uncertainties is 1.4\,\ms, while the estimated
intrinsic stellar jitter for HD\,159868 of $\sim$2\,\ms.)

We are in the process of developing a modified detection algorithm based on
an improved version of the Lomb-Scargle (LS) periodogram (Lomb 1976;
Scargle 1982). The standard LS technique estimates power in a
time-series of data by fitting sinusoids at fixed periods to generate
a periodogram. However, planets do not necessarily (indeed it would
seem  only infrequently) lie in the circular orbits which result in
sinusoidal Doppler variations. Ideally one would fit Keplerians,
however this requires generating a periodogram which is a function of
both period {\em and} eccentricity.

We have therefore generated what we call ``two dimensional Keplerian
Lomb-Scargle'' (2DKLS) periodograms, which are formed  by fitting
Keplerians at a grid of periods and eccentricities, and then
calculating power using Equation 7 of Cumming (2004). The period at
maximum power in the period-eccentricity plane, is then used as an
initial estimate for a non-linear least-squares Keplerian fit, this
time varying  all the Keplerian free parameters until the minimum reduced
$\chi^2$ ($\chi^2_{\nu}$) is determined. The uncertainties quoted in
this paper are derived from the diagonal terms in the covariance matrix, and 
only represent
a true estimation of the uncertainty on the fitted parameters in the
absence of degeneracy between those parameters. As a general rule,
Keplerian fits usually have some degeneracy between semi-amplitude
($K$) and eccentricity ($e$), resulting in the uncertainty estimates
on those terms being a lower limit.

\begin{figure}
\plotone{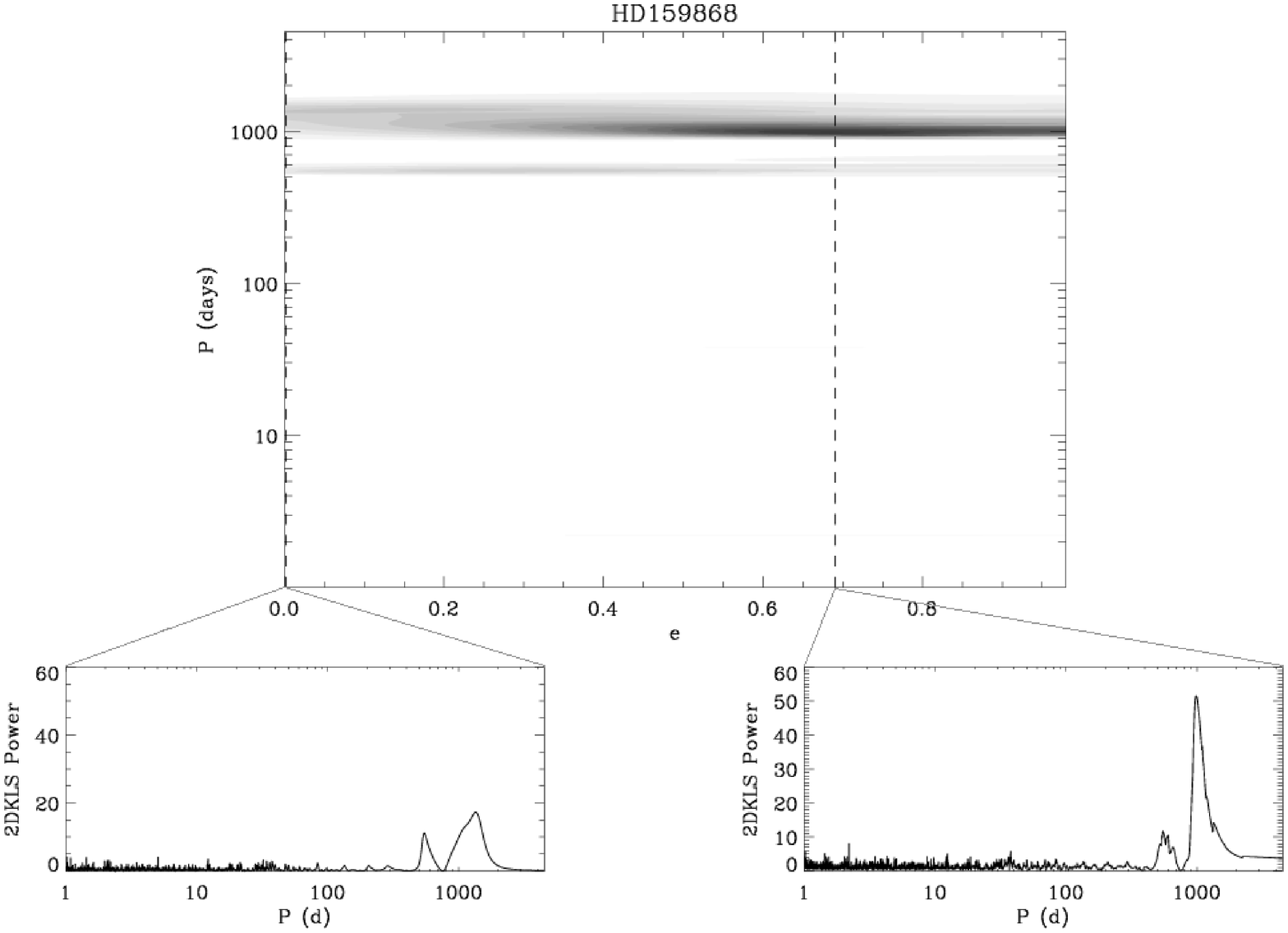}
\caption{2D Keplerian Lomb-Scargle (2DKLS) periodogram for HD\,159868.
The upper panel shows the 2DKLS, the lower left panel shows
the power spectrum generated at $e=0$, and the lower right panel
shows the power spectrum at $e=\pthreeecc$ corresponding to the peak
power in the 2DKLS. The power spectrum
at $e=\pthreeecc$ clearly shows higher contrast, and a different period
to that at $e=0$ (which corresponds to a ``standard'' Lomb-Scargle periodigram).}
\label{fig5}
\end{figure}

\begin{figure}
\plotone{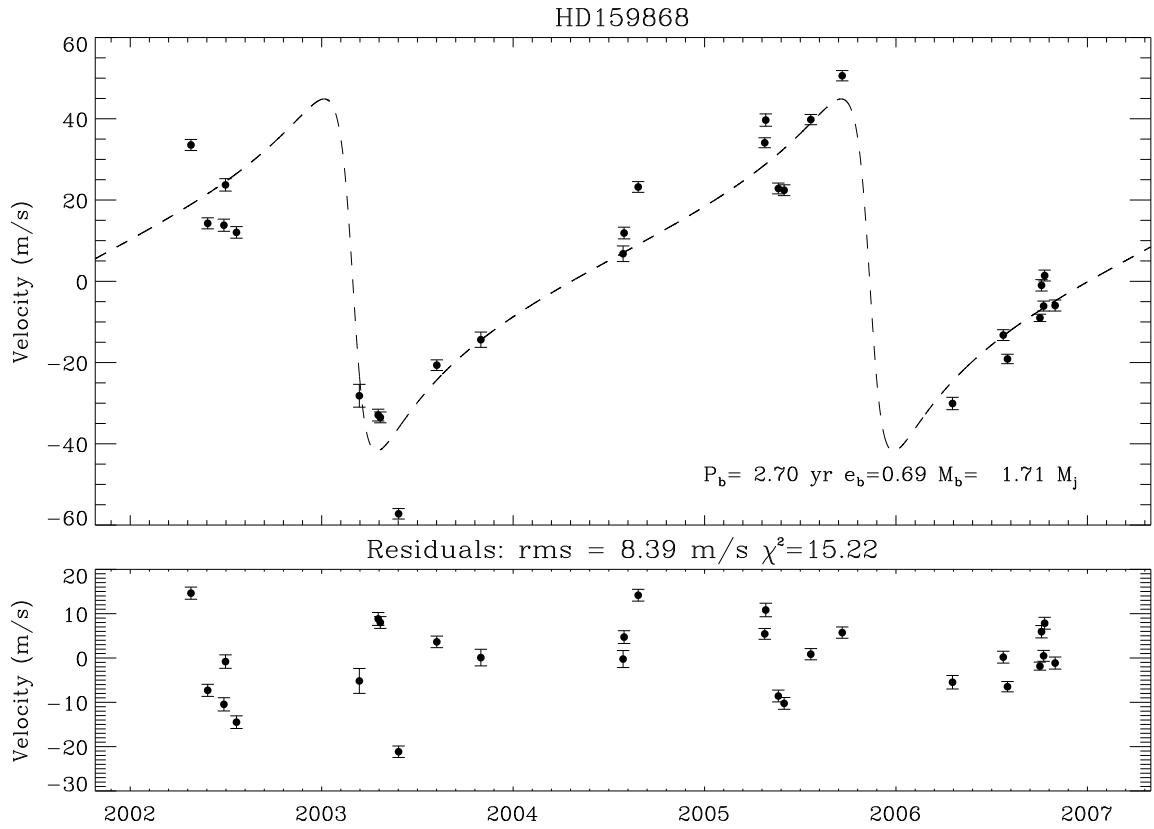}
\caption{Keplerian fit to HD\,159868. The RMS to a single Keplerian
  model is significantly higher than the internal measurement error,
  which combined with the high $\chi^2_{\nu}$ is suggestive of a second companion.}
\label{fig4}
\end{figure}

Figure \ref{fig5} demonstrates the 2DKLS with the Doppler data for
HD\,159868. The upper panel shows the 2DKLS power spectrum in greyscale, while the
lower panels show the power spectrum resulting at fixed eccentricities of
$e=0$ (left) and \pthreeecc\ (right). The former therefore corresponds
to a ``standard'' Lomb-Scargle periodogram (ie. fitting sinusoids),
and the latter is a cut at the eccentricity corresponding to peak
power in the 2DKLS.  The lower left panel of Figure \ref{fig5}
(i.e. for $e=0$) indicates maximum power at $P=1343$\,d, while that at
the 2DKLS peak power in the lower right panel is $P=986$\,d. The Keplerian
resulting from a non-linear least-squares solution is shown as a
dashed line in Figure \ref{fig4}, and has semi-amplitude
$K$=\,\pthreeamp\,\ms, giving a minimum planetary mass
\msini=\pthreemass\,\mjup, and a semimajor axis $a$=\pthreesep\,AU. 

Figure \ref{fig5} demonstrates some of the features 
of the 2DKLS method. The contrast in these power
spectra is considerably higher at the eccentricity corresponding to
peak 2DKLS power than at $e=0$. Moreover,
the period at which the power peak occurs is significantly different. Using the
standard LS as an initial estimate to the non-linear Keplerian fitting process
does eventually generate a best-fit period near 2DKLS peak power period.
However, using the intial estimate derived from the 2DKLS is clearly
preferable.

Throughout the remainder of this paper, when we present power spectra, these
will actually be cuts through the 2DKLS at the eccentricity
corresponding to the peak power in the 2DKLS (which is noted).

The  $\chi^2_{\nu}$ of the best-fit Keplerian is \pthreechisq, and the
RMS  is 8.5\,\ms, which is still significantly larger than expected
based on internal meaurement uncertainties and jitter.
Though a component of this excess RMS could be due to
asteroseismological ``noise'' (as discussed by Tinney et al.\ 2005),
it is unlikely to be the sole cause. So, while the Keplerian false
alarm probability (FAP -- calculated as described in Marcy et
al.\ 2005) for this fit is quite low ($<10^{-3}$), the large
residuals have motivated a search for additional companions.  

Allowing for a long term linear trend makes no significant improvement
to the RMS or $\chi^2_{\nu}$.  Planets with shorter periods can be fitted
simultaneously with the planet reported here, and solutions obtained that
significantly improve the RMS and $\chi^2_{\nu}$.  For example a
180\,d, \msini=0.5\,\mjup, $e$=0.05 planet reduces the joint RMS to
4.5\,\ms and $\chi^2_{\nu}$ to 4.4. However, the parameters of such a
solution are poorly constrained by the available sampling, so no
definitive statement can be made about the orbital properties of a
possible second planet around HD\,159868.

\subsection{HD\,23127}

HD\,23127 was added to the AAPS program in late 1998 together with 19
other targets, following suggestions that metal-rich stars 
preferentially host exoplanets (see e.g. Laughlin 2000 and
references therein). Tinney et al.\ (2003)  describes
this metal-rich subsample and its selection.  A total of
\ponenobs\ observations (listed in Table 4) have been
taken over 7.7 years beginning 1998 October. The RMS of these
velocities about the mean is 25\,\ms, again exceeding that expected
based on photon-counting internal measurement precision and stellar
jitter. (The median value of the internal measurement uncertainties is
4.4\,\ms; estimated  stellar jitter is $\sim$2.0\,\ms.)  Examination
of Table 4 shows how our precision  (as determined by the
internal uncertainty estimates listed in the table) has significantly
improved in recent years as exposure times have been increased from
$\sim$5 to $\sim$20 minutes to meet our higher S/N goal.

\begin{figure}
\plotone{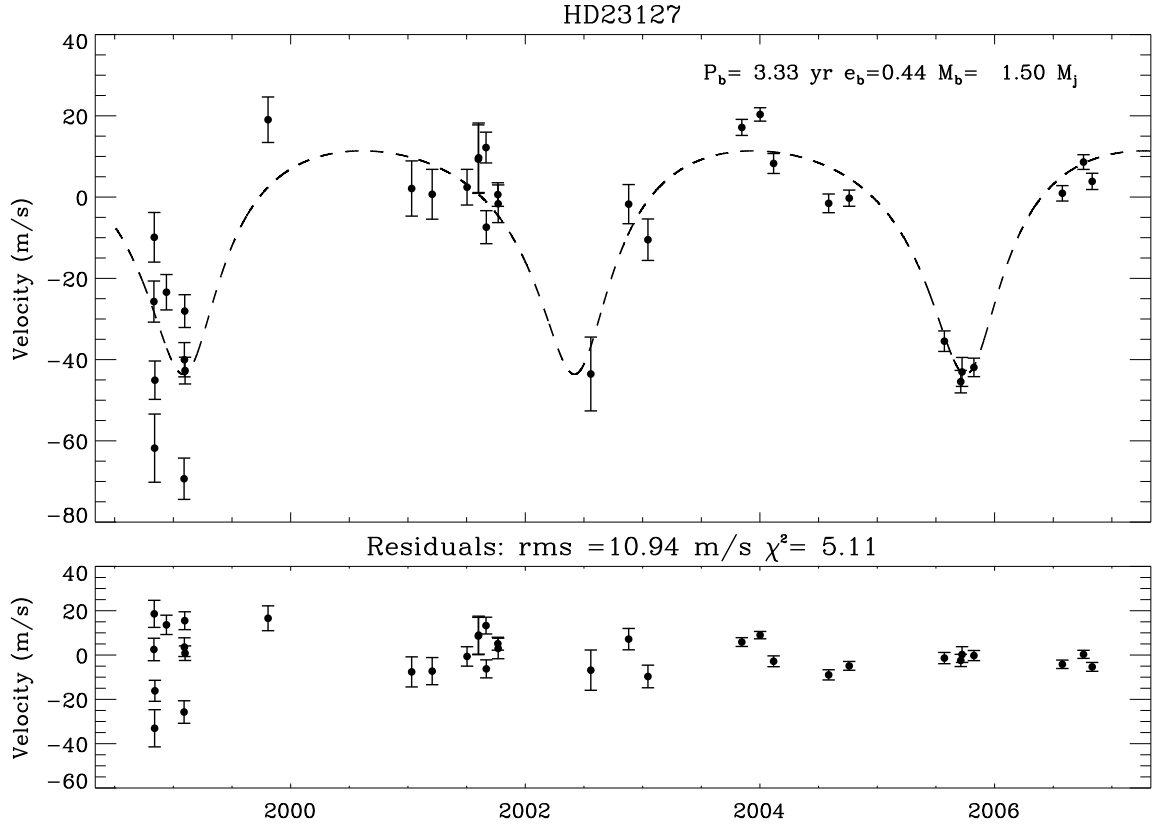}
\caption{Doppler velocities for HD\,23127, with the best--fit
  Keplerian (dashed line).  Most of the scatter in the Keplerian fit is due to
first epoch observations.  The observations taken over the first few
years have signal--to--noise of $\sim$50, compared to later observations
with S/N $\sim$100.}
\label{fig1}
\end{figure}

\begin{figure}
\plotone{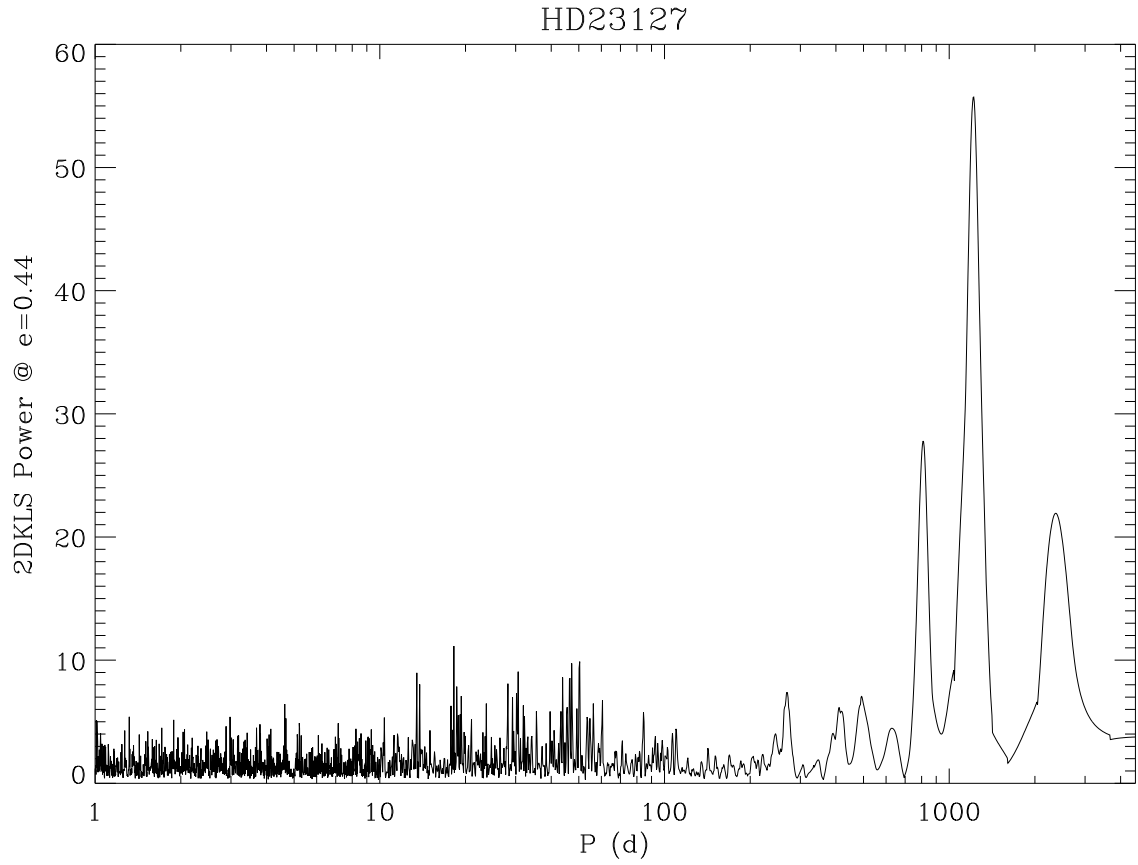}
\caption{2D Keplerian Lomb-Scargle periodogram for HD\,23127 at
  $e$=\poneecc\ showing significant power at 1214\,d or \poneper.
}
\label{fig2}
\end{figure}

The power spectrum for HD\,23127 is shown in Figure \ref{fig2} at an
eccentricity of \poneecc, corresponding to peak power in the
2DKLS. The peak at $\sim$1214\,d (\poneper)  is clear.  The best-fit
Keplerian is shown as the dashed line in Figure \ref{fig1}. The RMS of
the residuals (11.1\,\ms) and $\chi^2_\nu$ of the fit (\ponechisq) are
both high. These are due to the lower precision and larger scatter of
the earliest measurements and partially because of the star's
faintness. (If the residuals are divided in two, we find an RMS of
13.2\,\ms\ before 2003 and 4.8\,\ms\ after.) The minimum mass
(\msini) of the planet is \ponemass\,\mjup, and the semi-major axis is
\ponesep\,AU.

\subsection{HD\,154857}

HD\,154857 was added to our target list when the AAPS was
expanded in 2002.  As discussed in McCarthy et al.\
(2004) this star has evolved about 2 magnitudes above the
main sequence.  Based on 18 observations made between 2002 April
and 2004 February, McCarthy et al.\ announced a first planet
around this star, with a period of $\sim$400\,d, and an additional
linear trend suggesting a second longer period companion.

\begin{figure}
\plotone{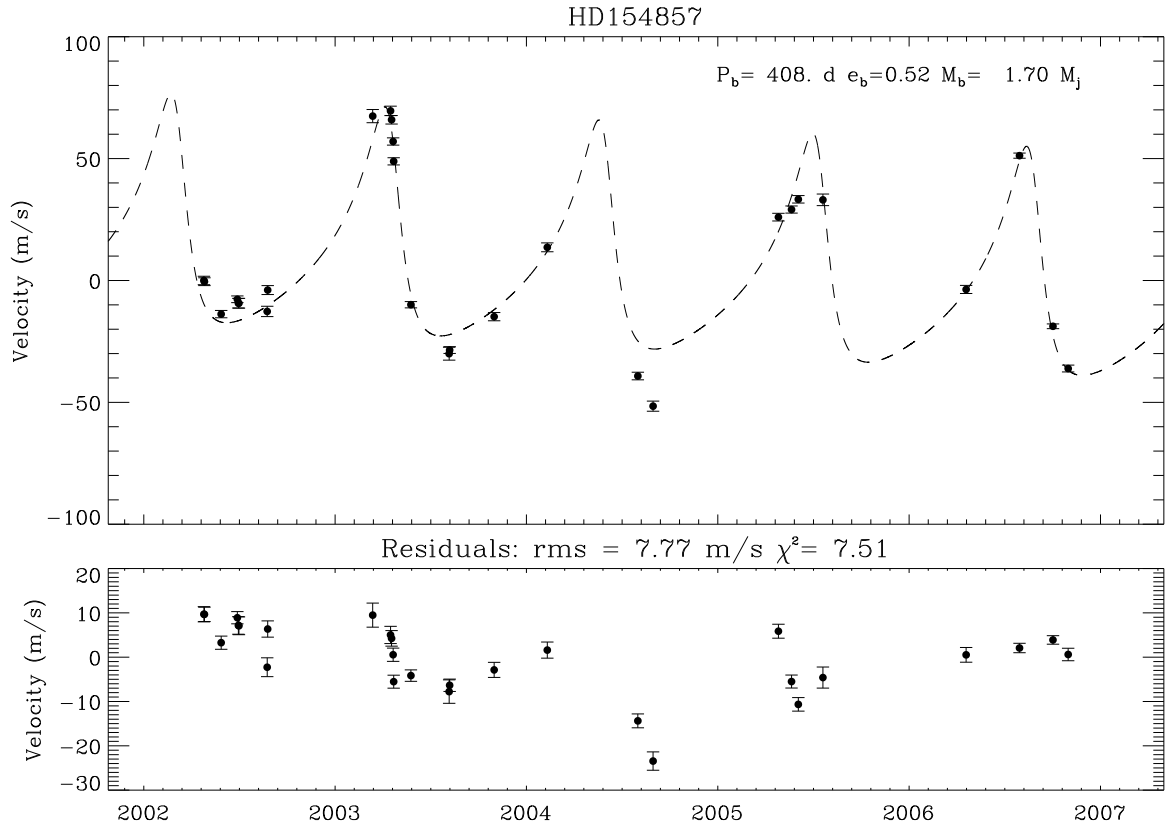}
\caption{A single Keplerian fit including a linear trend for
  HD\,154857. There is still a systematic variation present in the
  residuals and their RMS is still much higher than the internal measurement
  uncertainties.}
\label{fig7a}
\end{figure}

A total of \pfournobs\ AAT observations of HD\,154857, spanning 4.5
years, are listed in Table 5 and plotted in Figure
\ref{fig7a}. The velocity RMS of the raw data set is 35\,\ms.  The
median internal measurement uncertainty is 1.7\,\ms, and the estimated
stellar jitter is 2.6\,\ms.  The dashed line shows the best-fit single
Keplerian with a linear trend, with  an orbital period of 1.1\,yr and
an eccentricity of 0.52, in agreement with McCarthy et al. The RMS of
the velocities to the Keplerian plus linear trend is 7.77\,\ms\ and
the $\chi^2_{\nu}$ is 7.51.

\begin{figure}
\plotone{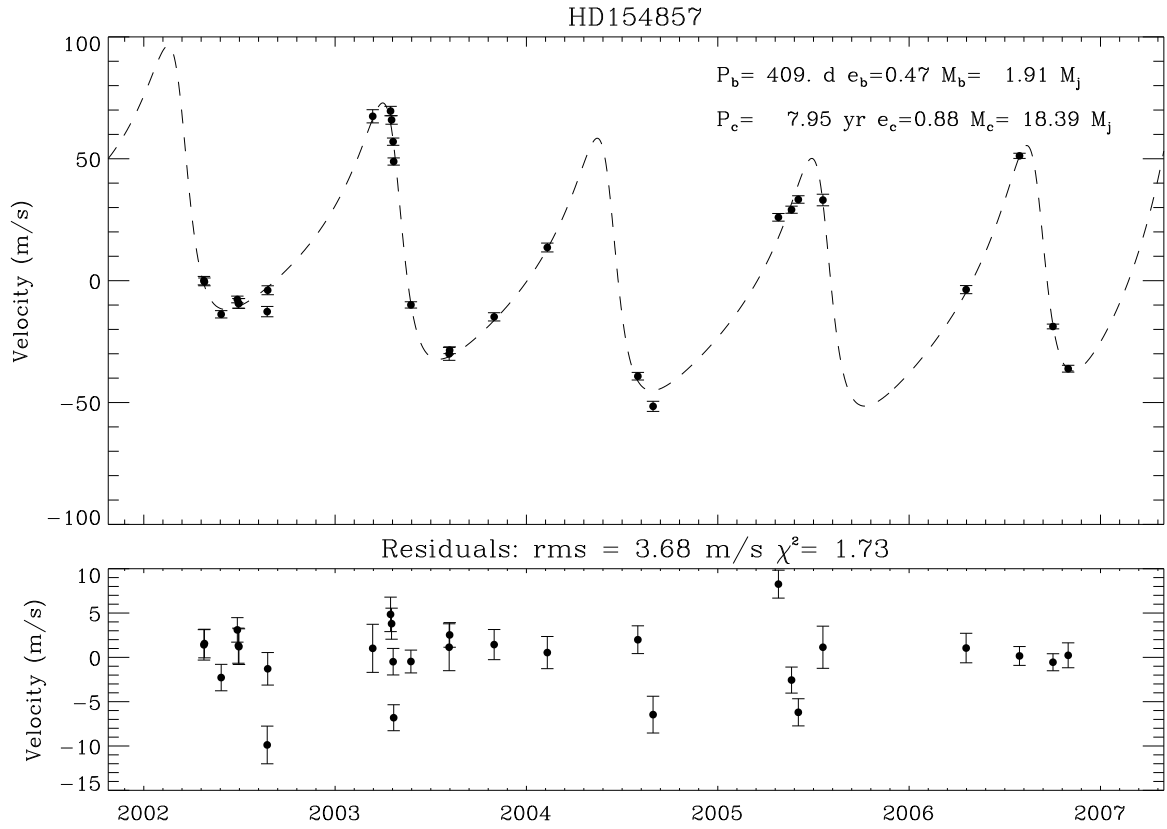}
\caption{A double Keplerian fit for HD\,154857. Fitting the second
  planet significantly improves the $\chi^2_{\nu}$, although its period and
  eccentricity are not well constrained.}
\label{fig7}
\end{figure}

Over the past two years the combination of a single Keplerian plus
linear trend has provided a systematically poorer match to the data.
We have therefore been motivated to search for a two Keplerian
solution.  Figure \ref{fig7} shows best-fit double Keplerian, which
was found by starting with the single-Keplerian parameters for the
known planet.  The updated orbital parameters for the inner planet
(Table 3) from this fit are basically consistent with those
given by McCarthy et al.\ (2004), and with four full orbits
observed, these parameters are now well constrained.  The fit parameters
for the outer planet are $P$=\pfourbper, $K$=\pfourbamp\,\ms, and
$e$=\pfourbecc.   The minimum mass is \msini=\pfourbmass\,\mjup\ and
$a$=\pfourbsep\,AU.  The RMS to the two Keplerian fit is 3.68\,\ms,
with a $\chi^2_{\nu}$ of \pfourchisq. 
The parameters of the (outer) long-period companion of
HD\,154857 are not well constrained; fitting very
long periods leads to poorly determined eccentricities and \msini\ values in
the brown dwarf regime or larger. More observations are needed to
constrain the parameters of this object, while robust lower limits are
provided in Table 3.

\section{Discussion}

Thirty planets have now emerged from the 260 target stars of the AAPS,
suggesting that $\sim$10\% of
late F, G, and K field dwarfs have planets that can be detected with
Doppler precisions of 3\,\ms\ and a time baseline of 8 years (a similar
detection rate to that of the original 106 stars on the Lick Observatory 
survey which has yielded 13 planets to date; Fischer et al.\ 2003).  These
surveys are now beginning to explore planets in orbits beyond 4\,AU,
though they remain insensitive to terrestrial mass planets beyond 0.1\,AU and
neptune-mass planets beyond 1\,AU. However, with planets being found orbiting
more than 10\% of nearby sun-like stars, it seems  that
planetary systems are common.

For the current high precision Doppler surveys, the one detectable
signpost of a Solar System analog is a giant planet in a circular orbit
($e<0.1$) beyond 4\,AU, with no giant planets interior to 4\,AU.
Previously five giant planets orbiting beyond 4\,AU have been found.
Of these, only 55 Cnc has an outer planet in such a 
circular orbit, however, it also has two giant planets
orbiting at 0.11 and 0.23\,AU.  The outer planet orbiting HD\,217107
(Vogt et al.\ 2005; Wittenmyer et al.\ 2006) has an eccentricity of
0.55 and an interior giant planet in a 7\,d orbit.
The planets orbiting HD\,72659 (Butler et al.\ 2006), HD\,154345 and
HD\,24040 (Wright et al. 2007) have
eccentricities of 0.27, 0.52, and 0.20 respectively. 

In this paper an additional object, possibly a giant planet (HD\,154857c) 
are presented, orbiting at (or beyond) 4\,AU and with quite
high eccentricity. Thus five of the six planets orbiting beyond 4\,AU,
are in eccentric orbits, and the sixth has giant planets in inner
orbits.  The distribution of eccentricities for planets beyond 4\,AU is
similar to that for planets between 0.2 and 4\,AU, and there is no
indication of a trend for increased circularity in planets orbiting
at large radii. This suggests that while gas-giant planets themselves
are common (orbiting $\simgt$10\% of sun-like stars), Solar System analogs 
may not be as common.  Over the
next decade the number of planets detected beyond 4\,AU will grow 
enabling us to determine accurately whether Solar System analogs are
rare or not.
However, to understand completely  the frequency of these detections,
we will need to more thoroughly simulate the selection functions
implicit in our observing strategies.
The AAPS program, along with other Doppler searches, has
been underway for more than 8 years. To date no detailed
simulation and analysis of these selection functions has yet been
undertaken by AAPS or any other planet search. What is needed is to
not only determine what can be detected using our current
observational techniques and sampling, but what should have been
detected but has not been, if we are to constrain the underlying
frequency distributions of planets and planetary system parameters.
An effort in this direction is now well underway for the AAPS.

\acknowledgements

We gratefully acknowledge the superb technical support at the
Anglo-Australian Telescope which has been critical to the success of
this project -- in particular we acknowledge R. Paterson,
D. Stafford, S. Lee, J. Pogson and J. Stevenson.  
We acknowledge support by PPARC grant PPC/C000552/1 (SJOT),
NSF grant AST-9988087 and
travel support from the Carnegie Institution of Washington (RPB),
NASA grant NAG5-8299 and NSF grant AST95-20443 (GWM).  
We thank the Australian and UK Telescope assignment
committees (ATAC \& PATT) for generous allocations of telescope time.  This
research has made use of NASA's Astrophysics Data System, and the SIMBAD
database, operated at CDS, Strasbourg, France.

\clearpage

\clearpage

\input{tab1.tex}

\clearpage

\input{tab2.tex}

\clearpage

\input{tab3.tex}

\clearpage

\input{tab4.tex}

\clearpage

\input{tab5.tex}

\end{document}

%% file: tab1.tex
\begin{deluxetable}{rrrllcclcccll}
\tablenum{1}
\tablecaption{Stellar Properties of Planet-bearing Stars\tablenotemark{a}}
\label{candidprop}
\tablewidth{0pt}
\tablehead{
\colhead{Star} & \colhead{Star} & \colhead{Spec.} & \colhead{B-V}
&\colhead{V} & \colhead{M$_\mathrm{V}$} & \colhead{Mass} & \colhead{\rhk} &\colhead{\vsini} & \colhead{Age} & \colhead{Jitter} & \colhead{[Fe/H]} & \colhead{d}
\\
\colhead{(HD)} & \colhead{(HIP)} &\colhead{Type} & \colhead{ }    & \colhead{} & \colhead{} & \colhead{(M$_{\odot}$)} & \colhead{ } & \colhead{(\kms)} & \colhead{(Gyr)} & \colhead{(\ms)} & \colhead{ } & \colhead{(pc)}
}
\startdata
 23127 &  17054 & G2V   & 0.69 & 8.58 & 3.82 & 1.13 & $-$5.00 & 3.3 & 7.1\rlap{\tablenotemark{b}} & 2.0 & +0.34 & 89.1 \\
% 92987 &  52472 & G2/3V & 0.64 & 7.03 & 3.81 & 1.08 & $-$5.01 & 2.5 & 8.2 & 1.8 & +0.05 & 44.0 \\
159868 &  86375 & G5V   & 0.72 & 7.24 & 3.07 & 1.09 & $-$4.96 & 2.1 & 8.1 & 2.0 & +0.00 & 52.7 \\
154857 &  84069 & G5V   & 0.65 & 7.24 & 3.63 & 1.17 & $-$5.00 & 1.4 & 5.6 & 2.6 & $-$0.22 & 68.5 \\
\enddata
\tablenotetext{a}{See Section \ref{sec:4new} for details on the
  references and derivations of the physical parameters in this
  Table. Typical uncertainties on these parameters are
  M$_\mathrm{V}\pm0.2$, mass\,$\pm0.10$, \rhk\,$\pm0.04$, \vsini\,$\pm0.5$, Age\,$\pm1.0$, Jitter\,$\pm0.5$, and [Fe/H]\,$\pm0.05$.}
\tablenotetext{b}{Lower limit for HD\,23127 is 4.2\,Gyr.}
\end{deluxetable}

%% file: tab2.tex
\begin{deluxetable}{rr}
\tablenum{2}
\tablecaption{Velocities for HD\,159868}
\label{vel159868}
\tablewidth{0pt}
\tablehead{
JD & RV \\
(-2450000)   &  (\ms) }
\startdata
%\tableline
  2390.2278  &    28.7 $\pm$ 1.3 \\
  2422.1471  &    10.5 $\pm$ 1.4 \\
  2453.0426  &    10.2 $\pm$ 1.5 \\
  2456.0712  &    19.6 $\pm$ 1.6 \\
  2477.0206  &     7.9 $\pm$ 1.5 \\
  2711.2689  &   -32.3 $\pm$ 2.7 \\
  2747.2526  &   -37.8 $\pm$ 1.4 \\
  2751.2604  &   -38.1 $\pm$ 1.3 \\
  2786.1000  &   -60.9 $\pm$ 1.3 \\
  2858.9541  &   -24.5 $\pm$ 1.3 \\
  2942.9349  &   -18.6 $\pm$ 1.9 \\
  3214.0987  &     2.9 $\pm$ 1.9 \\
  3216.0451  &     7.9 $\pm$ 1.4 \\
  3242.9683  &    19.2 $\pm$ 1.4 \\
  3484.2386  &    29.7 $\pm$ 1.2 \\
  3486.1651  &    35.3 $\pm$ 1.5 \\
  3510.1739  &    19.7 $\pm$ 1.3 \\
  3521.1956  &    18.4 $\pm$ 1.3 \\
  3572.0699  &    35.7 $\pm$ 1.2 \\
  3631.8981  &    46.5 $\pm$ 1.3 \\
  3842.2411  &   -34.0 $\pm$ 1.5 \\
  3939.0094  &   -17.4 $\pm$ 1.3 \\
  3947.0564  &   -23.5 $\pm$ 1.1 \\
  4008.9180  &    -9.0 $\pm$ 0.9 \\
  4011.9102  &    -1.0 $\pm$ 1.4 \\
  4015.9496  &    -6.1 $\pm$ 1.2 \\
  4017.8979  &     1.4 $\pm$ 1.3 \\
  4037.8961  &    -6.0 $\pm$ 1.4 \\
\enddata
\end{deluxetable}

%% file: tab3.tex
\begin{deluxetable}{llllllllll}
\tablenum{3}
\tablecaption{Orbital Parameters\tablenotemark{a}}
\label{candid}
\tablewidth{0pt}
\tablehead{
\colhead{Star}  & \colhead{Period} & \colhead{$K$} & \colhead{$e$} & \colhead{$\omega$} & \colhead{$T_0$} & \colhead{\msini} & \colhead{$a$} & \colhead{N$_{obs}$} & \colhead{RMS}
\\
\colhead{(HD)} & \colhead{(days)} & \colhead{(\ms)} &\colhead{ } & \colhead{($^\circ$)} & \colhead{(JD-2450000)}  & \colhead{(\mjup)} & {(AU)} & \colhead{ } & \colhead{(\ms)}
}
\startdata
 159868 &    986(9)  & \pthreeamp (2)  & \pthreeecc (0.02) &  97(3) &
 700(9)  & \pthreemass (0.3)  & \pthreesep (0.3) & \pthreenobs\ & 4.08 \\
  23127 &    1214(9)  & \poneamp (1)  & \poneecc (0.07) & 190(6) &
  229(19) & \ponemass (0.2) & \ponesep (0.3) & \ponenobs\ & 12.6 \\
   \\
%  92987\tablenotemark{a} &   3590(140)  & \ptwoamp (1)  & \ptwoecc (0.06) & 207(8) &
%  418(148) & \ptwomass (0.1) & \ptwosep (0.5) & \ptwonobs\ & 6.83 \\
%  92987\tablenotemark{b} &   3400:  & 21:  & 0.4: & &  & 1.5:  & 4.5:  & \ptwonobs\ & 6.02 \\
154857b &    409.0(1) & \pfouraamp (1)  & \pfouraecc (0.02) &  59(4)
& 346(5) & \pfouramass (0.4) & \pfourasep (0.2) & \pfournobs\ & 5.03 \\
%154857c\tablenotemark{a} &  2750(300) &  \pfourbamp (15)  & \pfourbecc
%(0.12) & 330(30) & 3430(400) & \pfourbmass (2.3) & \pfourbsep (0.8) & \pfournobs\ & 5.03 \\
154857c\tablenotemark{b} &  1900: &  23:  & 0.25: &  &  &  &  & \pfournobs\ & 5.03 \\

\enddata
\tablenotetext{a}{These values are the best-fit results from the 2DKLS
algorithm. $\omega$ and $T_0$ are the angle of periastron and
periastron passage time, respectively. Uncertainties are indicated in
parentheses, estimated as described in the text.}
\tablenotetext{b}{Numbers quoted with colons represent a robust lower
  limit; higher values produce a negligible difference in $\chi^2_{\nu}$.}
\end{deluxetable}

%% file: tab4.tex
\begin{deluxetable}{rr}
\tablenum{4}
\tablecaption{Velocities for HD\,23127}
\label{vel23127}
\tablewidth{0pt}
\tablehead{
JD & RV \\
(-2450000)   &  (\ms) }
\startdata
%\tableline
  1118.0928  &   -14.4 $\pm$ 5.1 \\
  1119.1769  &     1.5 $\pm$ 6.1 \\
  1120.2630  &   -50.4 $\pm$ 8.4 \\
  1121.1204  &   -33.7 $\pm$ 4.7 \\
  1157.1101  &   -12.0 $\pm$ 4.4 \\
  1211.9789  &   -58.0 $\pm$ 5.1 \\
  1212.9583  &   -28.7 $\pm$ 4.2 \\
  1213.9930  &   -16.7 $\pm$ 4.0 \\
  1214.9443  &   -31.3 $\pm$ 3.3 \\
  1473.2455  &    30.4 $\pm$ 5.6 \\
  1920.0076  &    13.5 $\pm$ 6.8 \\
  1983.8817  &    12.1 $\pm$ 6.1 \\
  2092.3152  &    13.8 $\pm$ 4.4 \\
  2127.2891  &    20.7 $\pm$ 8.4 \\
  2128.3130  &    21.0 $\pm$ 8.6 \\
  2151.3089  &    23.6 $\pm$ 3.8 \\
  2152.1997  &     4.0 $\pm$ 4.1 \\
  2188.1530  &    12.4 $\pm$ 2.9 \\
  2189.1692  &    12.3 $\pm$ 4.7 \\
  2477.3358  &   -32.3 $\pm$ 8.9 \\
  2595.0910  &     7.8 $\pm$ 4.8 \\
  2655.0337  &     0.3 $\pm$ 4.9 \\
  2947.1369  &    29.7 $\pm$ 2.0 \\
  3004.0254  &    31.7 $\pm$ 1.7 \\
  3045.9962  &    20.0 $\pm$ 2.4 \\
  3217.3038  &    10.1 $\pm$ 2.3 \\
  3281.2209  &    11.0 $\pm$ 2.0 \\
  3577.3179  &   -24.2 $\pm$ 2.5 \\
  3628.2861  &   -34.6 $\pm$ 2.7 \\
  3632.2557  &   -31.3 $\pm$ 3.5 \\
  3669.2013  &   -30.1 $\pm$ 2.3 \\
  3944.3322  &    12.0 $\pm$ 1.8 \\
  4010.1981  &     8.6 $\pm$ 1.8 \\
  4037.1442  &     3.9 $\pm$ 2.0 \\
\enddata
\end{deluxetable}

%% file: tab5.tex
\begin{deluxetable}{rr}
\tablenum{6}
\tablecaption{Velocities for HD\,154857}
\label{vel154857}
\tablewidth{0pt}
\tablehead{
JD & RV \\
(-2450000)   &  (\ms) }
\startdata
%\tableline
  2389.2358  &   -17.1 $\pm$ 1.7 \\
  2390.2122  &   -17.4 $\pm$ 1.6 \\
  2422.1371  &   -30.2 $\pm$ 1.5 \\
  2453.0201  &   -24.3 $\pm$ 1.4 \\
  2455.0253  &   -25.8 $\pm$ 2.0 \\
  2455.9766  &   -26.0 $\pm$ 2.0 \\
  2509.9485  &   -29.5 $\pm$ 2.1 \\
  2510.9162  &   -20.4 $\pm$ 1.9 \\
  2711.2461  &    50.4 $\pm$ 2.8 \\
  2745.2425  &    52.6 $\pm$ 1.9 \\
  2747.2118  &    48.8 $\pm$ 1.8 \\
  2750.1777  &    40.0 $\pm$ 1.5 \\
  2751.2295  &    31.8 $\pm$ 1.4 \\
  2784.1265  &   -26.9 $\pm$ 1.3 \\
  2857.0297  &   -47.0 $\pm$ 2.6 \\
  2857.9859  &   -45.6 $\pm$ 1.4 \\
  2942.9121  &   -31.9 $\pm$ 1.7 \\
  3044.2691  &    -3.1 $\pm$ 1.8 \\
  3217.0121  &   -56.0 $\pm$ 1.6 \\
  3246.0381  &   -68.0 $\pm$ 2.0 \\
  3485.1521  &     8.8 $\pm$ 1.6 \\
  3510.1595  &    12.3 $\pm$ 1.5 \\
  3523.1016  &    16.1 $\pm$ 1.6 \\
  3570.0292  &    16.0 $\pm$ 2.4 \\
  3843.2397  &   -20.8 $\pm$ 1.6 \\
  3945.0325  &    34.6 $\pm$ 1.1 \\
  4008.8962  &   -18.8 $\pm$ 1.0 \\
  4037.8808  &   -36.1 $\pm$ 1.4 \\
\enddata
\end{deluxetable}